\documentclass[a4paper,10pt,twoside]{cpc-hepnp}

\usepackage{multicol}
\usepackage{graphicx}
\usepackage{graphics}
\usepackage{booktabs}
\usepackage{amssymb,bm,mathrsfs,bbm,amscd}
\usepackage[tbtags]{amsmath}
\usepackage{lastpage}

\begin{document}

\fancyhead[c]{\small Submitted to Chinese Physics C} \fancyfoot[C]{\small 010201-\thepage}

\footnotetext[0]{Received 31 January 2014}

\title{Energy Spectrum of Cosmic Protons and Helium Nuclei by a Hybrid Measurement at 4300 m a.s.l.\thanks{Supported in China by NSFC (Contract
No. 10975145 and No. 11075170), the Knowledge Innovation Fund (H85451D0U2) of IHEP, the Chinese Ministry of Science and
Technology, the Chinese Academy of Science, and the Key
Laboratory of Particle Astrophysics, CAS,
and in Italy by
the Istituto Nazionale di Fisica Nucleare (INFN), and
Ministero dell'Istruzione, dell'Universit\`{a} e della Ricerca
(MIUR).}}
\author{%
\quad B.~Bartoli$^{2,3}$%
\quad P.~Bernardini$^{4,5}$
\quad X.J.~Bi$^{1}$
\quad I.~Bolognino$^{6,7}$
\quad P.~Branchini$^{8}$
\quad A.~Budano$^{8}$
\\\quad A.K.~Calabrese Melcarne$^{9}$
\quad P.~Camarri$^{10,11}$
\quad Z.~Cao$^{1}$
\quad R.~Cardarelli$^{11}$
% ...
\\(LHAASO Collaboration)\\
\\
}
\maketitle

\address{%
$^1$ Key Laboratory of Particle Astrophysics, Institute
                  of High Energy Physics, Chinese Academy of Sciences,
                  100049 Beijing, China.\\ 
$^2$ Dipartimento di Fisica dell'Universit\`a di Napoli
                  ``Federico II'', Complesso Universitario di Monte
                  Sant'Angelo, via Cinthia, 80126 Napoli, Italy.\\ 
$^3$ Istituto Nazionale di Fisica Nucleare, Sezione di
                  Napoli, Complesso Universitario di Monte
                  Sant'Angelo, via Cinthia, 80126 Napoli, Italy.\\ 
$^4$ Dipartimento di Matematica e Fisica "Ennio De Giorgi",
                  Universit\`a del Salento,
                  via per Arnesano, 73100 Lecce, Italy.\\ 
$^5$ Istituto Nazionale di Fisica Nucleare, Sezione di
                  Lecce, via per Arnesano, 73100 Lecce, Italy.\\ 
$^6$ Dipartimento di Fisica dell'Universit\`a di
                  Pavia, via Bassi 6, 27100 Pavia, Italy.\\
$^7$ Istituto Nazionale di Fisica Nucleare, Sezione di Pavia,
                  via Bassi 6, 27100 Pavia, Italy.\\
$^8$ Istituto Nazionale di Fisica Nucleare, Sezione di
                  Roma Tre, via della Vasca Navale 84, 00146 Roma, Italy.\\
$^9$ Istituto Nazionale di Fisica Nucleare - CNAF, Viale
                  Berti-Pichat 6/2, 40127 Bologna, Italy.\\
$^{10}$ Dipartimento di Fisica dell'Universit\`a di Roma ``Tor Vergata'',
                   via della Ricerca Scientifica 1, 00133 Roma, Italy.\\
$^{11}$ Istituto Nazionale di Fisica Nucleare, Sezione di
                   Roma Tor Vergata, via della Ricerca Scientifica 1,
                   00133 Roma, Italy.\\
$^{12}$ Tibet University, 850000 Lhasa, Xizang, China.\\ 
$^{13}$ Yunnan University, 2 North Cuihu Rd., 650091 Kunming,
                   Yunnan, China.\\ 
$^{14}$ Hebei Normal University, 050016 Shijiazhuang,
                   Hebei, China.\\
$^{15}$ Shandong University, 250100 Jinan, Shandong, China.\\
$^{16}$ Southwest Jiaotong University, 610031 Chengdu,
                   Sichuan, China.\\
$^{17}$ Dipartimento di Fisica dell'Universit\`a ``Roma Tre'',
                   via della Vasca Navale 84, 00146 Roma, Italy.\\
$^{18}$ Osservatorio Astrofisico di Torino dell'Istituto Nazionale
                   di Astrofisica, via P. Giuria 1, 10125 Torino, Italy.\\
$^{19}$ Istituto Nazionale di Fisica Nucleare,
                   Sezione di Torino, via P. Giuria 1, 10125 Torino, Italy.\\
$^{20}$ Dipartimento di Fisica dell'Universit\`a di
                   Torino, via P. Giuria 1, 10125 Torino, Italy.\\
$^{21}$ Physics Department, Kunming University, 650214 Kunming, Yunnan, China.\\
$^{22}$ Normal College of Fuyang, Fuyang 236029, China\\
}

\begin{abstract}
The energy spectrum of cosmic Hydrogen and Helium nuclei has been measured,
below the so-called ``knee", by using a hybrid experiment with a wide
field-of-view Cherenkov telescope and the Resistive Plate Chamber (RPC) array of the ARGO-YBJ
experiment at 4300 m above sea level.
The Hydrogen and Helium nuclei have been well separated from other cosmic ray
components by using a multi-parameter technique.
A highly uniform energy resolution of about $25\%$ is achieved
throughout the whole energy range (100~TeV - 700~TeV).
The observed energy spectrum is compatible with a single power law
with index $\gamma$=-2.63$\pm$0.06.
\end{abstract}

\begin{keyword}
Cherenkov telescope; ARGO-YBJ; energy spectrum; hybrid measurement; composition.
\end{keyword}

\begin{pacs}
98.70.Sa;96.50.sb;95.55.Cs
\end{pacs}

\footnotetext[0]{\hspace*{-3mm}\raisebox{0.3ex}{$\scriptstyle\copyright$}2013
Chinese Physical Society and the Institute of High Energy Physics
of the Chinese Academy of Sciences and the Institute
of Modern Physics of the Chinese Academy of Sciences and IOP Publishing Ltd}%

\begin{multicols}{2}

\section{Introduction}

The energy spectra of primary cosmic rays has been measured by many experiments around the ``knee".
However, all the experimental results do not precisely agree with each other~\cite{Horandel2003}.
 Convergence of controversial arguments about the origin of the ``knee" or more generally the origin of high energy cosmic rays has not been possible. This is due to the lack of a clean separation between species and of an independent energy scale determination in the experiments.
 Recently precise measurements
 have been carried out by the CREAM experiment which
 measured the energy spectra of individual nuclei with high statistical significance up to 50~TeV~\cite{CREAM}.
These measurements serve as standards for setting the energy scale and absolute abundance of each species.
At 4300 m a.s.l., the ARGO-YBJ experiment~\cite{ARGO-detector,ARGO-detector2,ARGO-detector3} is made of a fully covered array of single layer Resistive Plate Chambers (RPCs) in an area of 6700 m$^2$. The two unique features, namely the full coverage and the high altitude, enable the measurement to reach the lowest energy threshold with a ground-based detector. The energy spectrum of cosmic Hydrogen and Helium nuclei from 5~TeV to 200~TeV has been measured by the ARGO-YBJ experiment~\cite{argo-spectrum} and agrees well with the sum of the two individual components measured by the CREAM experiment~\cite{CREAM}.
Moreover, the ARGO-YBJ equipped two channels of large dynamic range analog readout for the total charge in each RPC. This allows to measure high energy showers up to a few PeV without saturation of the detector. The charge measurement of the RPCs maps the detailed particle distribution in the shower core region.
   Two prototype telescopes of the Wide Field-of-view (FOV) Cherenkov
 Telescope Array (WFCTA)~\cite{WFCTA_telescope}, a component of the future project Large High Altitude Air Shower Array (LHAASO)~\cite{LHAASO-caozh,LHAASO-hhh}, are deployed to the Yangbajing International Cosmic Ray Observatory in Tibet to form a hybrid experiment together with the RPC array. The combination of the two air shower detecting techniques enhances the selection sensitivity for specific composition of primary cosmic rays and the robustness of the primary energy measurements.  In this paper, the hybrid observation, the selection of cosmic Hydrogen and Helium nuclei, the energy measurement and the energy spectrum of the selected  $H$ and $He$ nuclei are reported.
\section{The Experiment}
The hybrid experiment with the two telescopes and the ARGO-YBJ RPC array located at the Yangbajing
Cosmic Ray Laboratory (Tibet, People¡¯s Republic of China, 4300 m a.s.l.)
started the air shower observations in 2008.
One of Cherenkov telescope, named WFCT-01, is located 99.1 m away from the center of the RPC array outside the north-west corner of the ARGO experiment hall. The other one, named WFCT-02, is located at the south-east corner of the ARGO-YBJ detector, about 78.9~m away from
the center of the RPC array.
Each telescope, equipped with 16$\times$16 photomultipliers (PMTs), has a FOV of $14^{\circ}\times16^{\circ}$ with a pixel size of approximately $1^{\circ}$~\cite{WFCTA_telescope}. The telescopes are tilted up with the main optical axes having an elevation of 60$^\circ$ and image showers coming in the FOV around 30$^\circ$ with respect to the zenith.
The ARGO-YBJ $78\times74 ~m^2$ central carpet consists of 1836 RPCs covering ~93\% active area~\cite{argo10, argo-RPC, argo11, ARGO-detector-2, argo-sat}.
Each chamber is equipped with two channels of analog
to collect the total charge induced
by particles passing through each half of the chamber, called ``big pad"~\cite{big-pad, bidpad-michele} with a size of 139~$\times$~123~cm$^2$. The total charge is proportional to the number of charged particles
~\cite{bidpad-michele, bidpad-mxh, lixxnimpaper}.

From December 2010 to February 2012, the coincident cosmic ray events that trigger both WFCT-02 and the RPC array simultaneously
are used in the data analysis.
The main constraint on the exposure of the hybrid experiment is the weather condition in the moon-less operational nights. The weather is monitored in two ways. At first, the bright stars in the FOV of the telescope are used as constant light sources to monitor the  total optical depth~\cite{llma-star}.
The  telescope measures the background light intensity using
the DC coupled PMT signals as the baseline. It
varies as bright stars pass
through the FOV of the PMT. The variation is clearly correlated with the total stellar flux which  can be obtained from the star catalog~\cite{star-catalog}, as long as the sky is clean.
In other words, the correlation coefficient serves as an indicator of the weather condition.
Secondly, an infrared camera covering the whole sky monitors the clouds above the horizon.
Confirmed by the infrared monitor, a correlation coefficient greater than 0.8 defines good weather conditions~\cite{llma-star} suitable for the Cherenkov imaging of showers.
Combining the good weather selection and the live time of the RPC array, the total exposure time is 7.28$\times10^5$ seconds for the hybrid measurement.

Criteria for well measured showers are 1) their cores must be located inside the ARGO-YBJ
carpet excluding an edge of 2 meters; 2) at least 1000 hits recorded by the RPCs in order to have high quality geometrical reconstruction of the shower fronts for demanded angular
 and core position resolutions~\cite{ARGO-resolution}; 3)
at least 6 triggered pixels in each shower image are required and 4) the space angle between the incident direction of
the showers and the telescope main axis, denoted as $\alpha$, must be less than 6$^\circ$ to guarantee that the images are fully contained in the FOV.
About 32,700 events survived  the cuts and were well reconstructed in the aperture of the hybrid experiment, namely zenith angle from $24^{\circ}$ to $37^{\circ}$ and
azimuth angle from $249^{\circ}$ to $273^{\circ}$ and core within an area of the RPC array of 76~m $\times$ 72~m.

\begin{center}
\includegraphics[width=9cm]{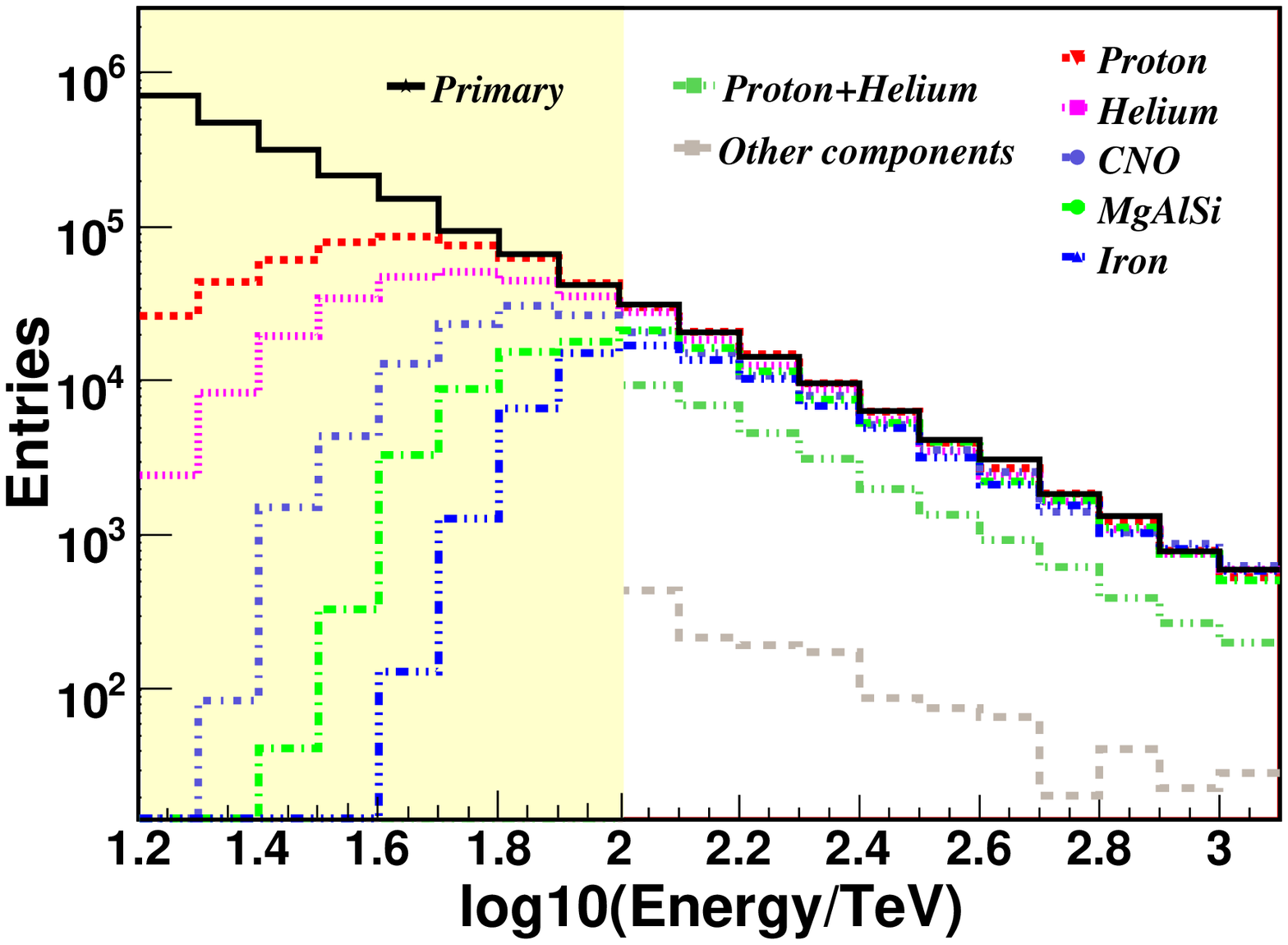}
\figcaption{\label{energe-cut}
Simulated energy distributions of the five primary particle groups (protons, helium, CNO, MgAlSi and iron) after the quality cuts (see Sect. 2) in the hybrid experiment. It is clearly shown that the detector almost reaches a full efficiency for all species above 100 TeV. The energy distributions of the light nuclei and of the other components after the selection (see Sect. 4) are shown by dashed curves initiating at 100 TeV. The injected primary energy spectrum (solid line) is also shown for reference.}
\end{center}
\begin{center}
\includegraphics[width=9cm]{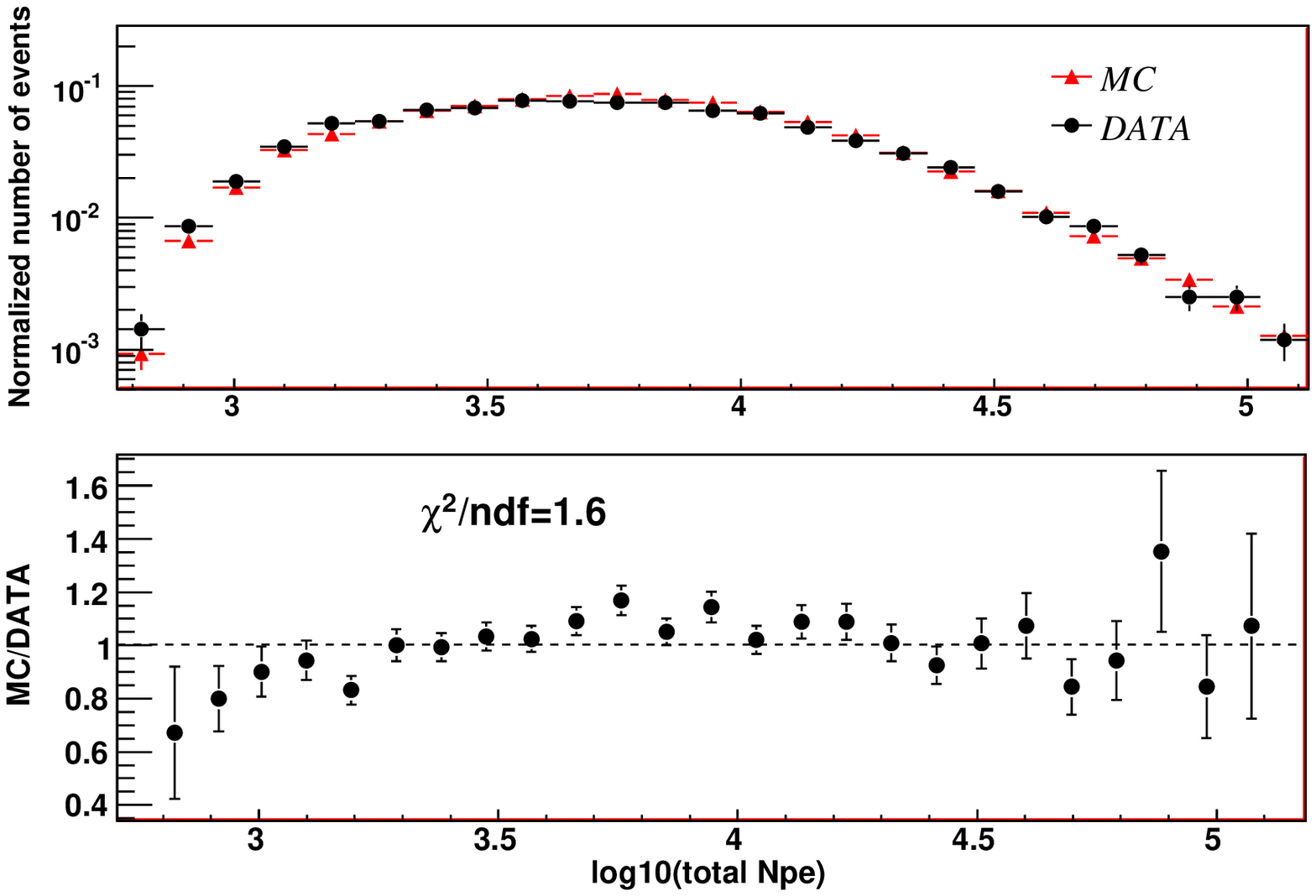}
\figcaption{\label{Npe-distribution}
 The distributions of the total number of photoelectrons $N_{pe}$ in shower images.
 In the upper panel the dots represent the observed events and the triangles the simulated ones.
 In the lower panel the ratio between the simulated and the observed number of events is shown.}
\end{center}
\section{Simulation}
Extensive air shower simulations including Cherenkov photons are carried out by using the CORSIKA code~\cite{corsika}
with the high energy hadronic interaction model QGSJETII-03~\cite{QGSJET} and the low energy model
GHEISHA~\cite{GHEISHA}.
A GEANT4-based simulation package (G4argo)~\cite{G4argo} is used for the ARGO-YBJ
detector. A ray tracing procedure on the Cherenkov photons~\cite{CRTNT} is also carried out of the response of Cherenkov telescopes.
In the simulation, the primary particles are divided into five groups: proton, helium, CNO (carbon, nitrogen and oxygen) group,
MgAlSi (magnesium, alumina and silicon) group and iron in the simulation. Assuming a spectral index of -2.70 for all the  five groups up to 10 PeV,
the energy distributions of simulated showers in the individual groups are shown in Fig.$\ref{energe-cut}$ by using the same selection criteria as described before for the data. We point out that
the hybrid observation becomes nearly full efficient above 100~TeV for all components.
To test the simulation, a comparison with data has been made by means of the distributions of total number of photoelectrons in the shower images, zenith angles of the shower arrival directions and impact parameter of the showers.
For this comparison, a more realistic composition and spectral model given by Horandel~\cite{Horandel2003} is assumed in the simulation. The results are presented in Fig.$\ref{Npe-distribution}$,Fig.$\ref{theta-distribution}$ and Fig.$\ref{Rp-distribution}$.
The Monte Carlo simulation represents the data reasonably well according to the $\chi^2$ per degree of freedom.

\begin{center}
\includegraphics[width=9cm]{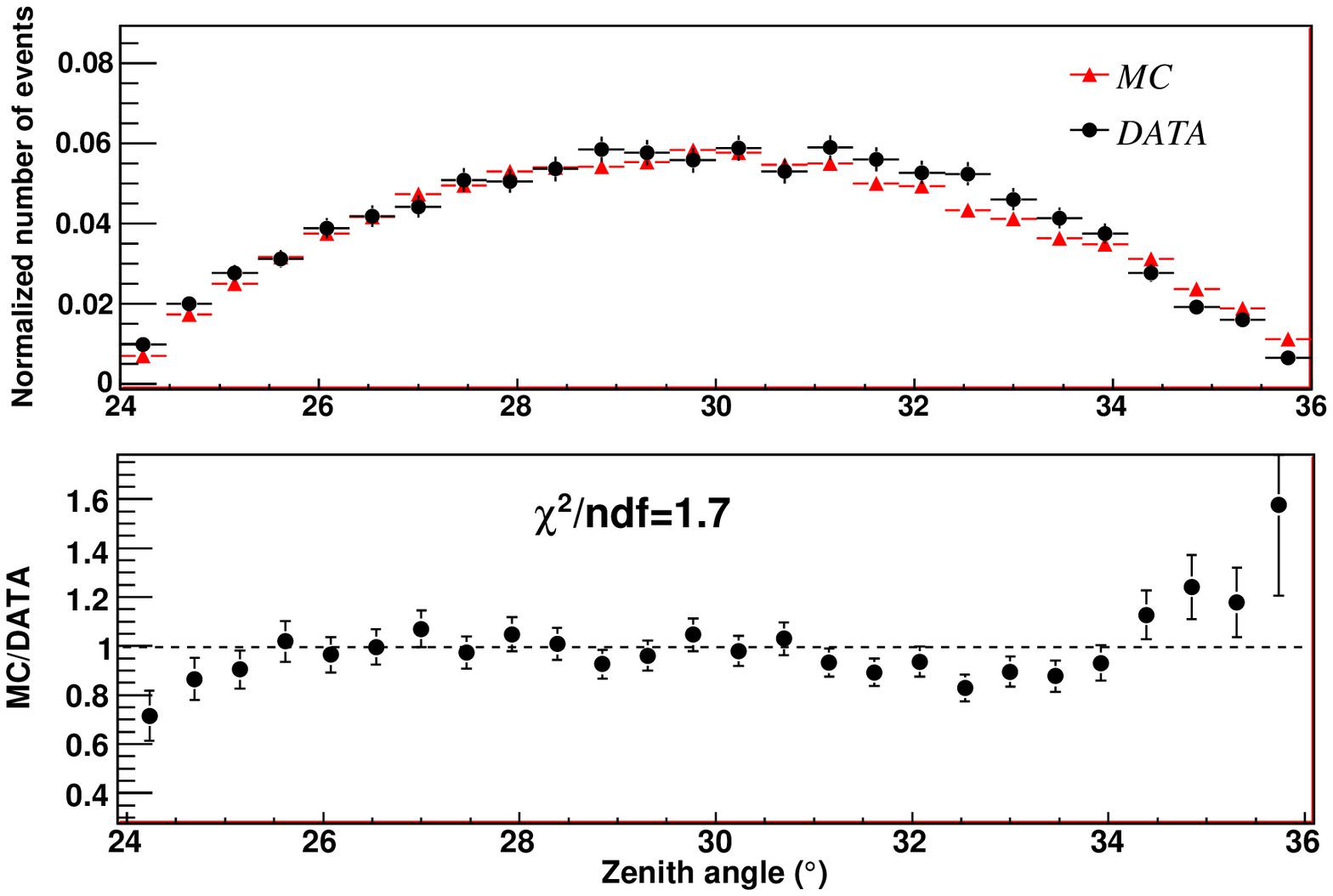}
\figcaption{\label{theta-distribution} The comparison of the zenith angle distributions for simulated and observed events. Symbols are as in Fig.$\ref{Npe-distribution}$}
\end{center}

\begin{center}
\includegraphics[width=9cm]{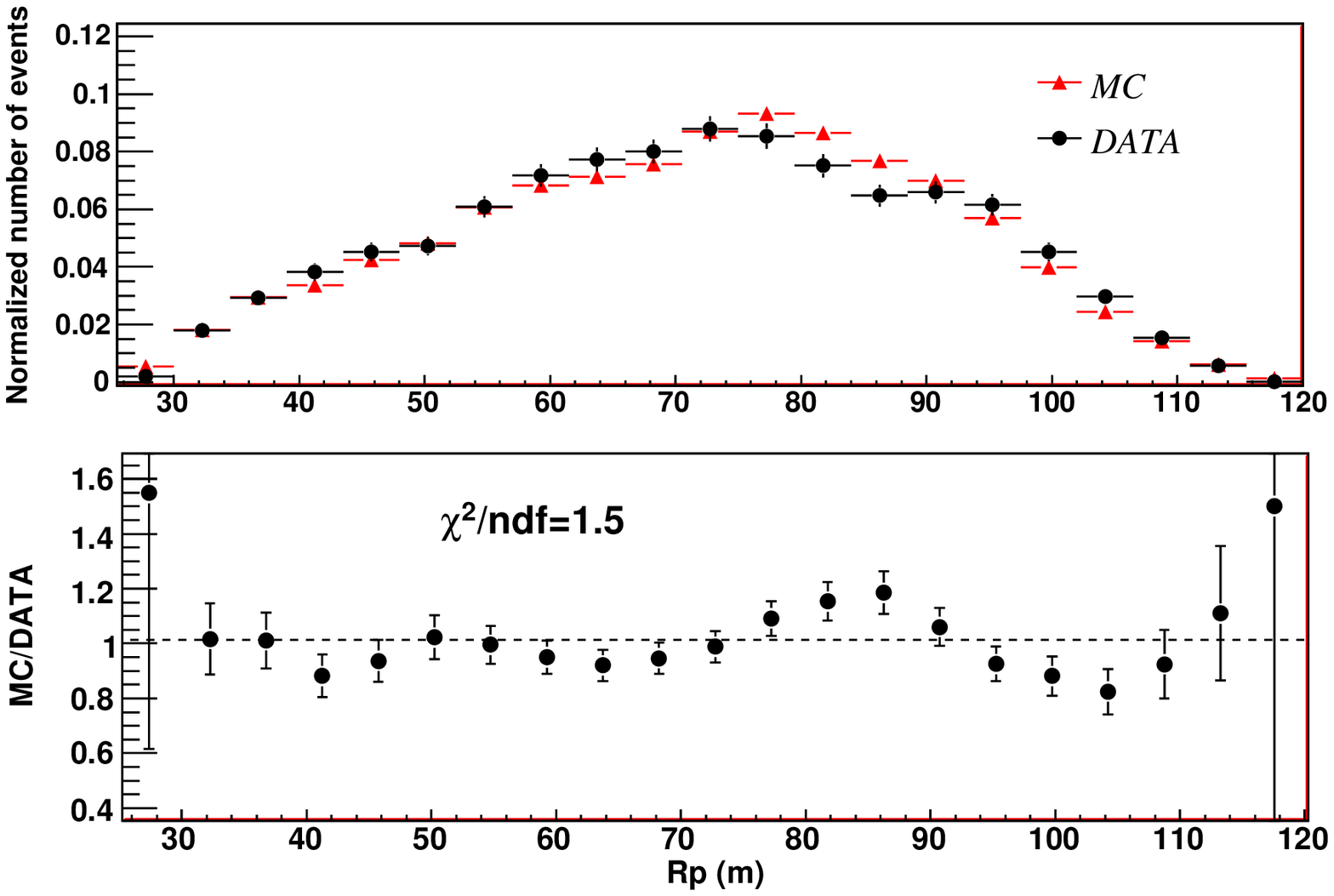}
\figcaption{\label{Rp-distribution} The comparison of the impact parameter distributions for simulated and observed events. Symbols are as in Fig.$\ref{Npe-distribution}$}
\end{center}

Based on this simulation, an investigation aiming at the selection of Hydrogen and Helium induced showers, out of all detected showers, cosmic ray showers is carried out as follows.

\section{Hydrogen and Helium Nuclei Selection}
The secondary particles
in showers induced by heavy nuclei are further spread away from the core region where a uniform lateral distribution
due to Coulomb scattering is well described by the Nishimura-Kamata-Greisen (NKG) function. Therefore, it is clearly seen that there exist
significant differences between the lateral distributions around the core of showers induced by light and heavy nuclei,
while they are similar beyond a certain distance, e.g. 20~m.
With its full coverage, the ARGO-YBJ array uniquely measures the lateral distribution of secondary particle densities near the shower cores.
Usually the largest number of particles recorded by a RPC in an event, denoted as $N_{max}$, is a good measure of the lateral distribution in 3 meters from the core. In a shower induced by a heavy nucleus $N_{max}$ is expected smaller than that
in a shower induced by a light nucleus with the same energy.
According to the simulation, $N_{max}$ is also proportional to $E_{rec}^{1.44}$, where $E_{rec}$
is the reconstructed primary energy using the Cherenkov telescope (see Sect. 5) as the first order approximation, without knowing the composition of the shower.
The reduced parameter $log_{10} N_{max}$ - 1.44$log_{10}(E_{rec}/1TeV)$, denoted as $p_L$, is a good indicator of the nature of the primary.
As an example, the separation between the proton and iron showers is a factor of 2 on average.

The other mass sensitive parameter is associated with the shape of the Cherenkov images of showers recorded by the telescope.
 The elliptic Cherenkov image of a shower is described by the Hillas parameters~\cite{hillas}, such as the width and length of the image.
The images are more stretched, i.e. narrower and longer, for showers that are more deeply developed in the atmosphere. The ratio of the length to the width ($L/W$) is therefore a parameter
sensitive to the primary composition.
It is also known that the images are more elongated for showers farther away from the telescope due to pure geometrical reasons. This effect can be removed by using the well measured shower impact parameters, $R_p$. Moreover,
the images are also more stretched for more energetic showers due to the elongation of the cascade processes in the atmosphere. This effect can be suppressed by using the ``energy" $E_{rec}$. According to simulations,
the ratio $L/W$ of images is linearly proportional to $R_p$ and $log_{10}E_{rec}$. The reduced parameter
$L/W-0.0091\times(R_p/1m)-0.14log_{10}(E_{rec}/1TeV)$, denoted as $p_C$, serves as an indicator of the nature of the primary that initiated the shower. As an example, the separation between the proton and iron showers is a factor of 1.5 on average.

Combining the two composition-sensitive parameters, $p_L$ and $p_C$,
one expects the separation between cosmic ray components will be improved.
This is shown in Fig.$\ref{Rpcmax-L-W}$ where all the simulated events are displayed in a scatter plot of the two parameters.
Protons, helium, CNO group, MgAlSi group and iron with the ratio of 1:1:1:1:1 are put in the simulation. At first, no strong correlation between the two parameters is observed, indicating that the parameters are quite independent. Secondly, a rather significant separation between the composition groups is clearly observed, although the different groups overlap each other. Thirdly, the lighter components, e.g. $H$ and $He$, are in the uppermost-right region while the iron showers are mainly concentrated in the lower-left corner. Finally, it is rather significant that the fluctuation in showers initiated by heavier nuclei is much less than that in showers induced by light nuclei. This offers a great opportunity to pick out a light composition sample with high purity by simply cutting off the concentrated heavy cluster in the lower-left region in the scatter plot, i.e.
  $p_L\leq -0.91$ and  $p_C\leq 1.3$.
Most of the heavy nuclei (CNO group, MgAlSi group and iron) are cut out
with a contamination less than 5.1$\%$ among the survived $H$ and $He$ samples.
This contamination reduces to 2.3\% if a more realistic composition model is assumed as the Horandel model~\cite{Horandel2003}.
 About 29.7$\%$ of $H$ and $He$ survives the selection criteria and their energy distribution
is shown in Fig.$\ref{energe-cut}$ as the nearly parallel but lower curve. The small portion of remaining heavy nuclei is also shown in the figure as the lowest curve.

\begin{center}
\includegraphics[width=9cm]{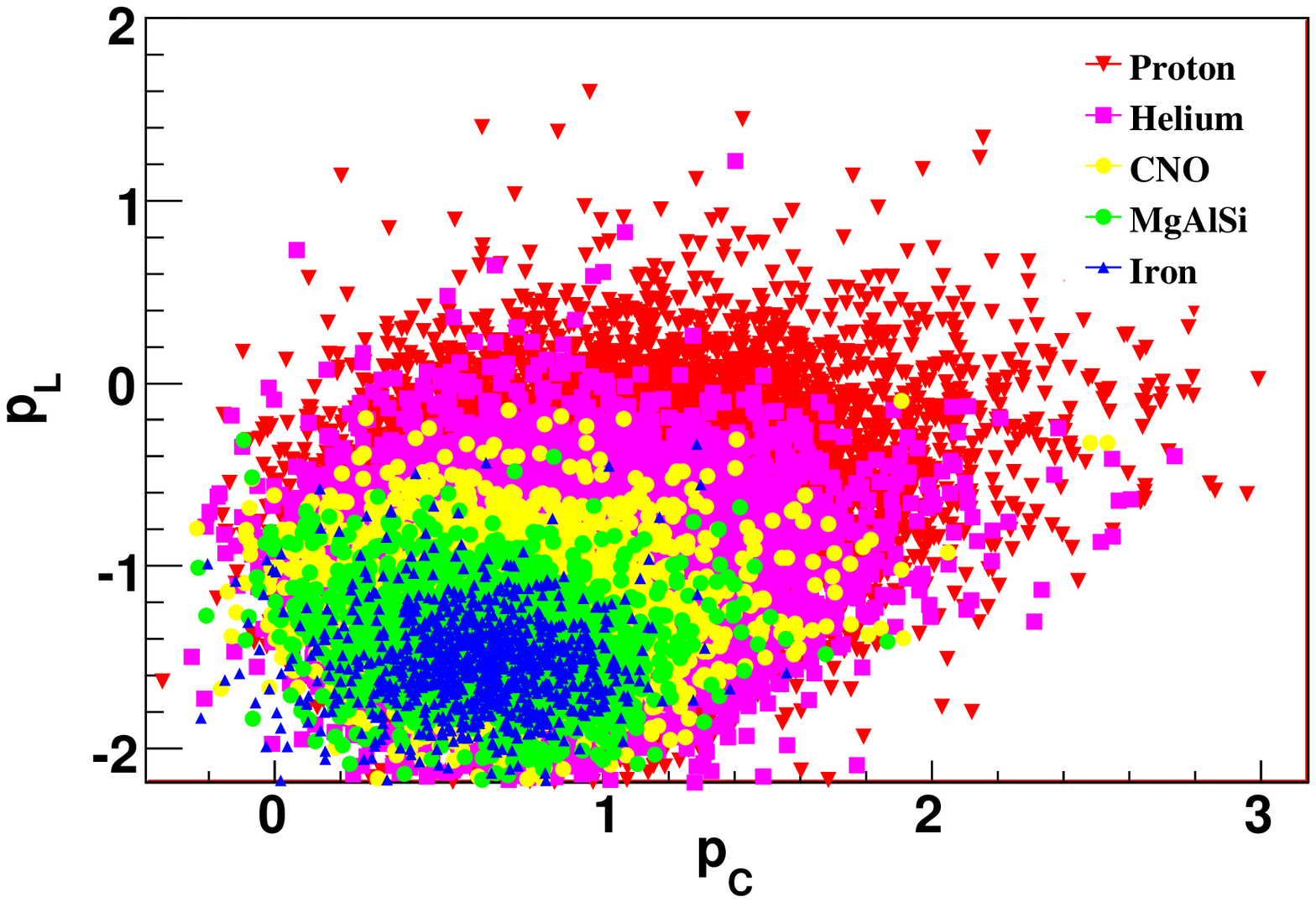}
\figcaption{\label{Rpcmax-L-W}Scatter plot of the two mass-sensitive parameters $p_L$ and $p_C$ (see text for details).}
\end{center}

As mentioned above, the hybrid experiment is almost fully efficient to all showers above 100 TeV. The aperture is estimated using
the Horandel model for the primary composition and the QGSJET/GHEISHA code to describe the hadronic interaction.
It is shown as filled circles in Fig.$\ref{aperture}$. It is approximately a constant of 170~$m^{2}sr$  above 100~TeV. The aperture of the $H$ and $He$ detection using the hybrid experiment shrinks to 50.5~$m^{2}sr$ above 100~TeV by taking into account the selection efficiency. The aperture remains
constant with energy as also shown in the figure. No extra bias is introduced by the $H$ and $He$ selection.

The systematic uncertainty on the aperture can be estimated by modifying the composition assumed in the simulation, e.g. the CREAM measurement results~\cite{CREAM}
 or extreme cases such as the heavy nuclei dominant model or the proton dominant model~\cite{HD}.
The effect on the selection efficiency is not greater than $14.3\%$.
The contamination by heavier nuclei is quite stable, from $5.1\%$ to $2.3\%$ as the composition assumption changes from one extreme to the other. Using the SIBYLL code, instead of QGSJET, the selection efficiency is found to be about 2.3$\%$ higher. The difference in the efficiency due to the low-energy hadronic interaction models, GHEISHA or FLUKA, is about $3.5\%$. The overall uncertainty on the aperture is 14.9\%, as shown in Fig.\ref{aperture}.
\begin{center}
\centering
\includegraphics[width=9cm]{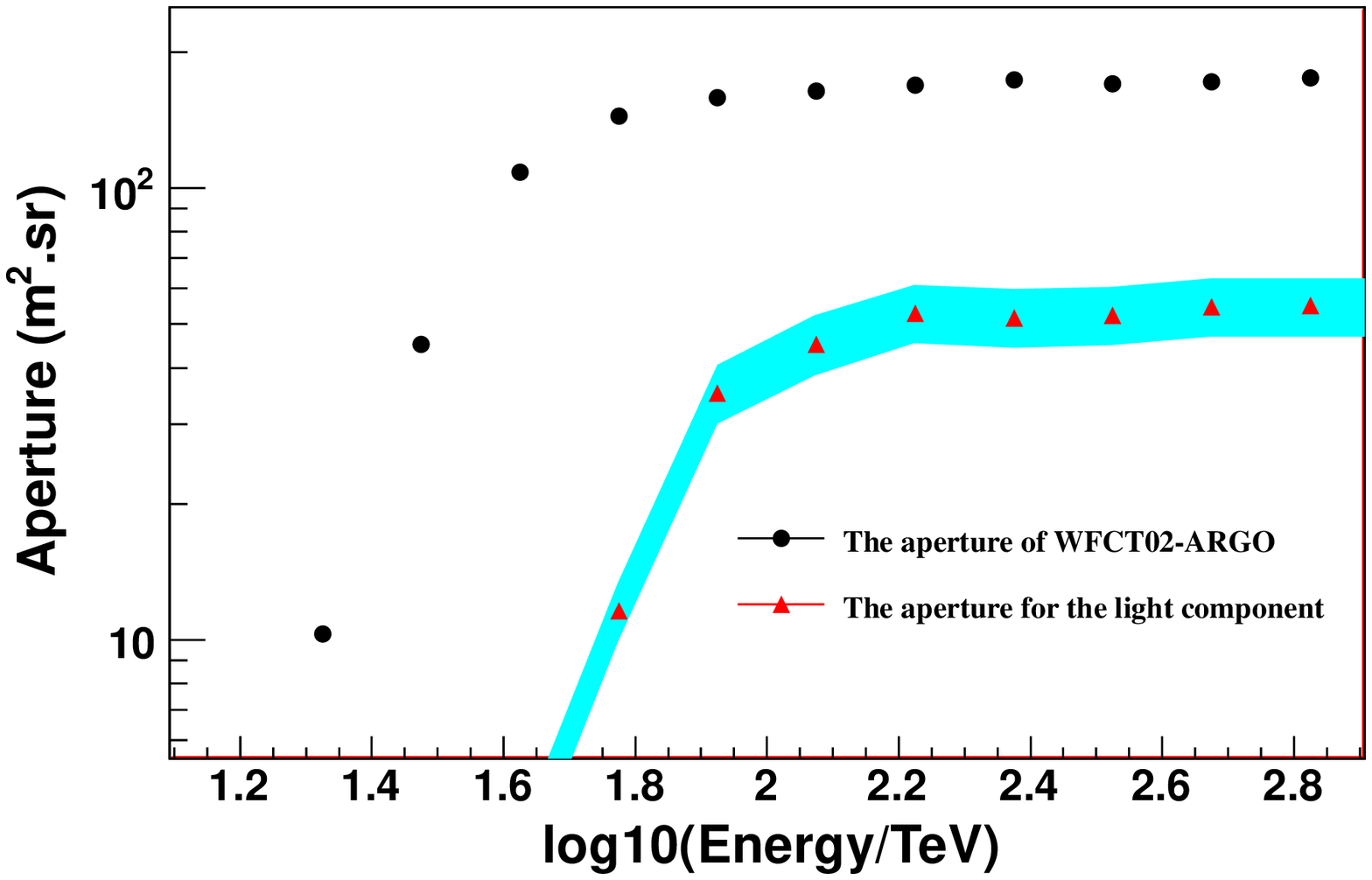}
\figcaption{\label{aperture}
The aperture of the hybrid experiment. Filled circles are for all particles, triangles are for selected protons and helium nuclei. The shaded area around the triangles represents the systematic uncertainty.}
\end{center}

\section{Energy measurement}

The energy of the primary cosmic ray initiating the shower is estimated by using the total number of photoelectrons, $N_{pe}$, collected in the image recorded by the telescope, which results from all Cherenkov photons
produced in the whole history of the shower development. For selected showers falling in the RPC array, the telescopes is at distances shorter than 120~m from the shower cores. The $N_{pe}$ varies dramatically with the impact parameters, $R_p$, because of the rapid falling off of the lateral distribution of the Cherenkov light.
An accurate determination of the shower geometry is crucial for the energy measurement.
The angular and core resolutions of the geometrical reconstruction using the RPC array are better than 0.4$^\circ$ and 2~m, respectively.
In the FOV of $14^\circ\times 16^\circ$ of the telescope,
the $N_{pe}$ still varies slightly with the incident angle, $\alpha$.
A look-up table established by using the simulation is a way to reconstruct the energy of the primary for such a complicated functional form. By feeding in the three measured variables $N_{pe}$, $R_p$ and $\alpha$, the primary energy can be interpolated in the table. For the selected $H$ and $He$ sample, the table is generated with a mixture of only protons and helium nuclei.
The energy resolution is about $25\%$ and is symmetric and uniform from 100 TeV up to a few PeV.
The systematic bias is less than $2\%$ throughout the entire energy range.
The intrinsic fluctuation of the shower development is the main contribution to
the energy resolution.
However, there is also a contribution from the primary mixing, the resolution being about 21\% for a pure proton sample.

The systematic uncertainty in the energy reconstruction is mainly due to the following items. 1) The uncertainty due to the composition assumed in the simulation is estimated by switching between the three models mentioned above. It turns out to be very small, i.e. about 1.2$\%$ in
the energy scale.
2) The uncertainty due to the hadronic interaction models adopted in the simulation ( QGSJET or SYBILL, and GHEISHA or FLUKA ) is found to be less than $2.0\%$.
3) The uncertainty due to the photometric calibration, which has an uncertainty of 7\%, is estimated to be $5.6\%$.
More details about the absolute calibration can be found elsewhere~\cite{WFCTA_telescope}. 4) The uncertainty due to the weather condition is estimated by using the
starlight of the Galactic plane recorded by the telescope. A variance $<9.5\%$ in the light intensity is observed after the good weather selection. This corresponds to an energy underestimate of about $7.6\%$.
In total, the overall systematic uncertainty in
the energy scale is $\sim$~$9.7\%$.
\section{Results and Discussion}
Applying the criteria mentioned in Sect. 2 on the data set taken by the WFCT-02 and ARGO-YBJ hybrid experiment, 8218 events above 100 TeV are selected.
They are distributed in the $E_{rec}$-$p_C$-$p_L$ space as shown in Fig.$\ref{Scatter-plot}$, where $E_{rec}$ is the estimated energy of the primary. From this sample, 1392 $H$ and $He$ like shower events are selected.
The energy distribution of these events is shown in Table\ref{table-flux} and the statistics error are smaller than 20\% in each bin.
To take into account any kind of smearing and migration from the true energy E of the primary to the reconstructed energy $E_{rec}$, the Bayesian  method~\cite{bayesian} is used to unfold the observational data.

\begin{center}
\includegraphics[width=9cm]{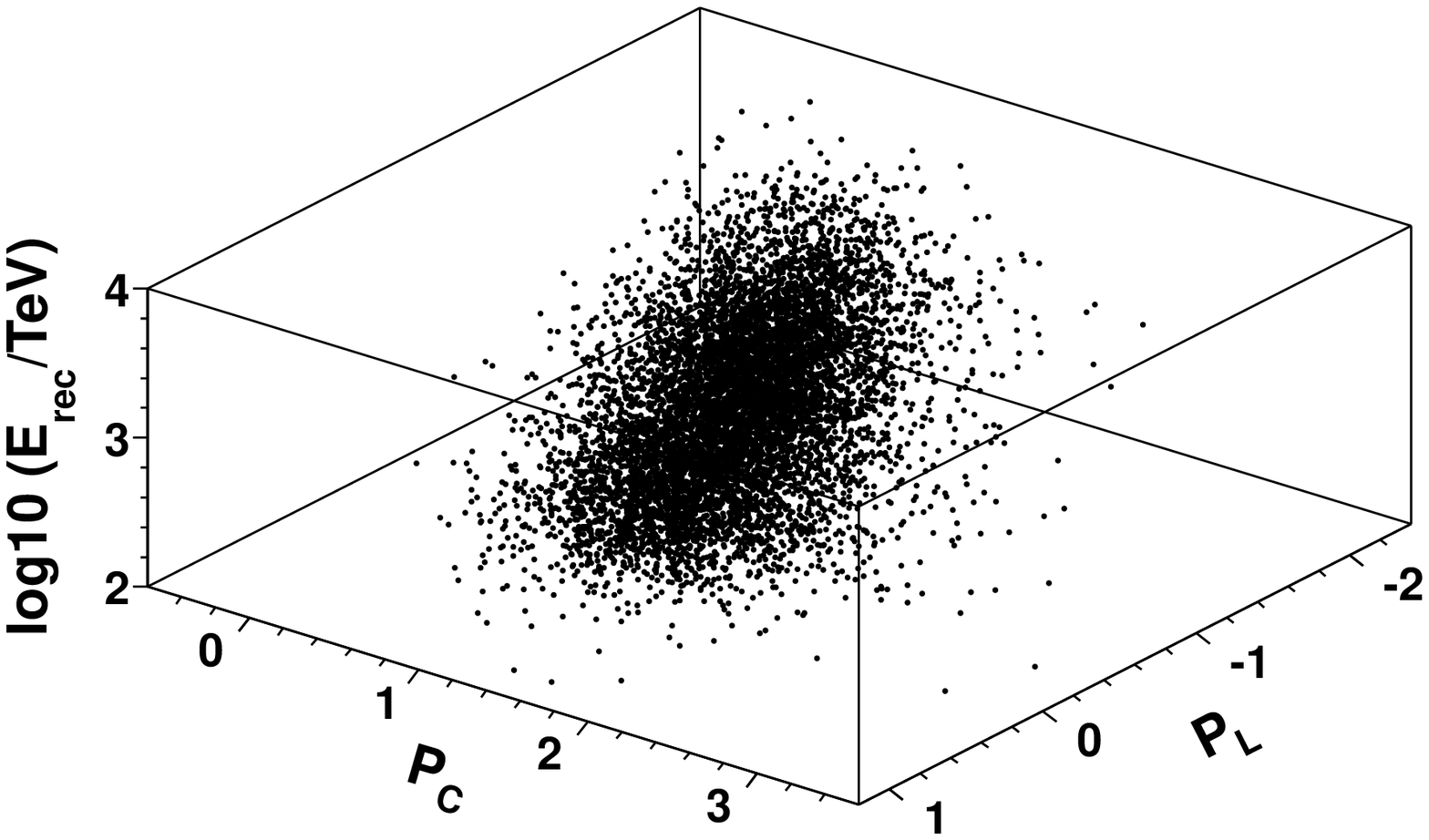}
\figcaption{\label{Scatter-plot}The scatter plot for all the selected events in the $E_{rec}$-$p_C$-$p_L$ space.}
\end{center}

The energy spectrum of the cosmic Hydrogen an Helium nuclei measured by the hybrid experiment
is shown in Fig.$\ref{Spectrum-result}$ as the filled squares.
A power law with a single spectral index of -2.63 $\pm$ 0.06 fits the spectrum well and the $\chi^2/ndf$ is about 0.5. The absolute flux at 400~TeV is $(1.79\pm0.16)\times 10^{-11}~GeV^{-1}~m^{-2}~sr^{-1}~s^{-1}$.
The systematic uncertainty in the absolute flux is 14.9\% as indicated by the shaded area around the squares in Fig.$\ref{Spectrum-result}$. The error bars are statistical only.

\end{multicols}
\begin{center}
\tabcaption{\label{table-flux}The number of protons and helium like events in each energy bin.}
\footnotesize
\begin{tabular*}{170mm}{@{\extracolsep{\fill}}ccccccc}
 \toprule Log(E/1TeV) bins\hphantom{00} &\hphantom{0} 2.00-2.15 & \hphantom{0}2.15-2.30  & \hphantom{0}2.30-2.45 & \hphantom{0}2.45-2.60 &\hphantom{0} 2.60-2.75  &\hphantom{0} 2.75-2.90 \\
   \# of events \hphantom{00}&\hphantom{0}565 & \hphantom{0}371 & \hphantom{0}227 & \hphantom{0}121 & \hphantom{0}69 & \hphantom{0}39 \\
\bottomrule
\end{tabular*}
\end{center}
\begin{multicols}{2}

This result almost fills the energy gap between the measurements above 1 PeV , such as those by the KASCADE experiment~\cite{Kascade-results},
and the spectrum of Hydrogen and Helium nuclei measured up to 200 TeV by the ARGO-YBJ experiment~\cite{argo-spectrum}.
The latter is consistent with the new
 measurement using the hybrid technique in the overlapping energy region from 100 to 200~TeV. The flux difference is less than 10\%.
The spectrum by the ARGO-YBJ alone is important because it reaches a much lower energy, 5~TeV, therefore overlaps the CREAM spectrum~\cite{CREAM}, which is measured by a calorimeter calibrated with an Indium beam at 158~GeV/nucleon or 18~TeV/particle~\cite{CREAM_calibration} and with a proton beam at 350 GeV/nucleon~\cite{CREAM_calibration2}. The consistency between the two measurements within 10\% in the overlapping energy region
guarantees that the energy scale difference between the two experiments is less than 4\%.
This is important for the combination of all the three independent measurements covering a wide energy region from 2~TeV to 700~TeV.
The sum of proton and helium spectra measured by CREAM~\cite{CREAM} is fitted by using a power law with a single spectral index of -2.62$\pm$0.02. The index of  -2.61$\pm$0.04 is reported by the ARGO-YBJ experiment~\cite{argo-spectrum}. Combining them together with -2.63$\pm$0.06 as the result of the hybrid experiment, there is no strong evidence of any structure of the spectrum of cosmic protons and helium nuclei up to 700~TeV.

\begin{center}
\includegraphics[width=9cm]{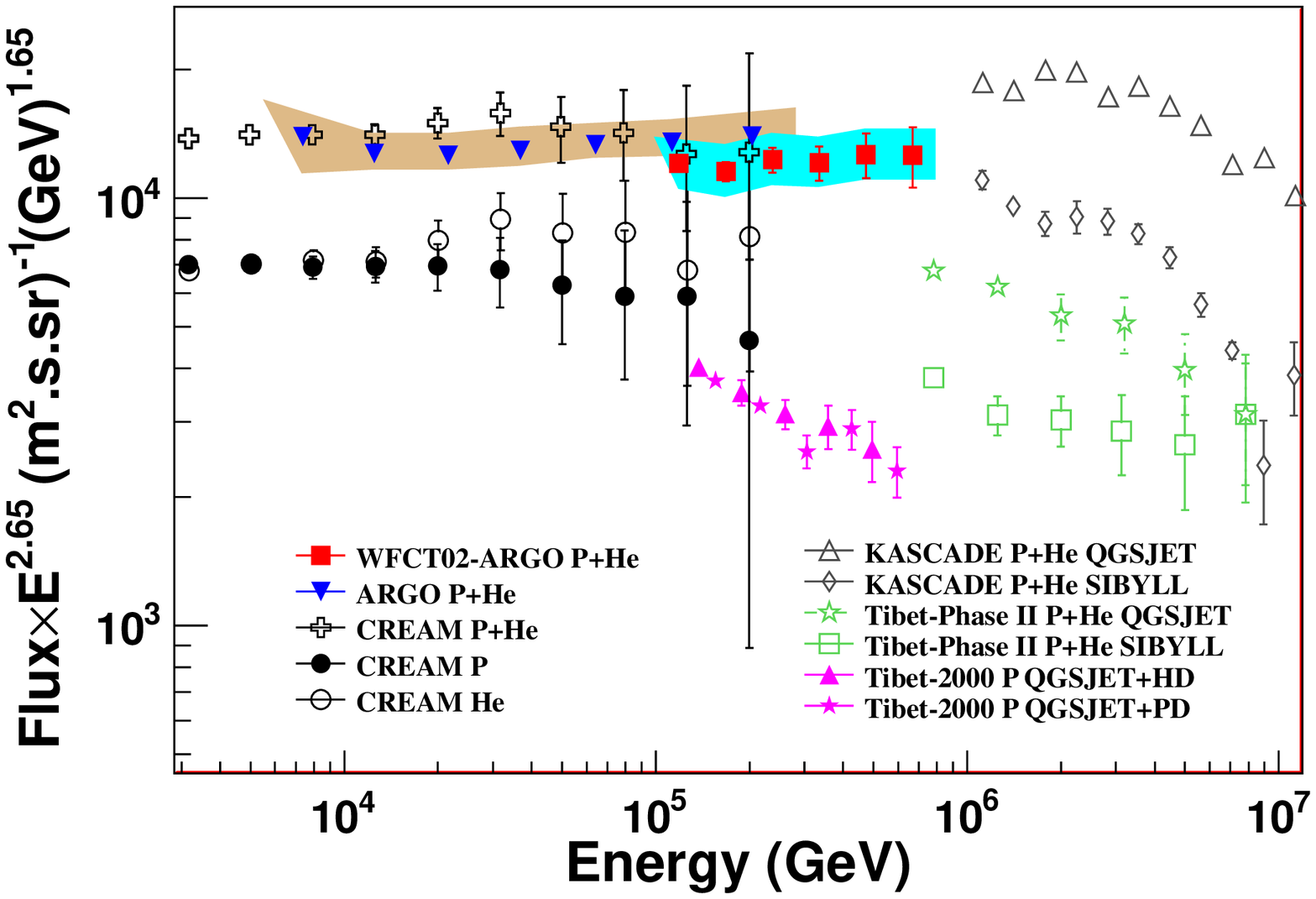}
\figcaption{\label{Spectrum-result}
 The spectrum of cosmic protons and helium nuclei from 100 TeV to 700 TeV measured by the hybrid experiment is shown by filled squares. For comparison, the CREAM spectra of protons, helium nuclei and their sum are shown by filled and open circles and open crosses , respectively. The ARGO-YBJ results are represented by inverted triangles. Systematic uncertainties are indicated by the shaded areas for both the hybrid and the ARGO-YBJ experiments. Other ground-based experimental results are also plotted for comparison.}
\end{center}

 In a similar energy range, from 200~TeV to
1~PeV, the Tibet air-shower experiment obtained the energy
spectrum of pure protons, and pure helium nuclei~\cite{ASgama-knee}.
By estimating the spectral index to be around -2.97$\pm$0.06,
the Tibet air-shower experiment claimed that the proton spectrum was probably being bent at an energy of around 100 TeV
if the measured spectrum has to be smoothly connected with the existing direct measurements at lower energies, such as CREAM.

 In summary, the energy spectrum of cosmic protons and helium nuclei from 100 to 700~TeV is measured by the hybrid experiment using the Cherenkov telescope WFCT-02 and the RPC array of the ARGO-YBJ experiment. The overall systematic uncertainty in the absolute flux is smaller than 14.9\%. The uncertainty in energy determination is about 9.7\%.
This measurement agrees in both spectral index and absolute flux with the spectrum obtained by ARGO-YBJ alone in the lower energy range from 5 TeV to 200 TeV.
 The latter agrees with the CREAM measurements within 4\% in the energy scale, so the energy scale of this measurement is confirmed.
The current measurement extends the spectrum up to 700~TeV.
In conclusion, no significant structure deviating from a power law with a single index is found in the energy spectrum of the light component from 5~TeV to 700~TeV.

\acknowledgments{We acknowledge the essential support of
W.Y. Chen, G. Yang, X. F. Yuan, C.Y. Zhao, R. Assiro,
B. Biondo, S. Bricola, F. Budano, A. Corvaglia, B.
D'Aquino, R. Esposito, A. Innocente, A. Mangano, E.
Pastori, C. Pinto, E. Reali, F. Taurino, and A. Zerbini, in
the installation, debugging, and maintenance of the
detector.}

\end{multicols}

\vspace{10mm}

\vspace{-1mm}
\centerline{\rule{80mm}{0.1pt}}
\vspace{2mm}

\begin{multicols}{2}

\end{multicols}

\clearpage

\end{document}